\documentclass[conference]{IEEEtran}
\IEEEoverridecommandlockouts
\usepackage{cite}
\usepackage{amsmath,amssymb,amsfonts}
\usepackage{algorithmic}
\usepackage{graphicx}
\usepackage{textcomp}
\usepackage{xcolor}
\usepackage{soul}
\usepackage{url}
\usepackage[caption=false]{subfig}
\usepackage{gensymb}

\makeatletter % changes the catcode of @ to 11
\newcommand{\linebreakand}{%
  \end{@IEEEauthorhalign}
  \hfill\mbox{}%\par
  \mbox{}\hfill\begin{@IEEEauthorhalign}
}
\makeatother % changes the catcode of @ back to 12

\def\BibTeX{{\rm B\kern-.05em{\sc i\kern-.025em b}\kern-.08em
    T\kern-.1667em\lower.7ex\hbox{E}\kern-.125emX}}
\begin{document}

\title{Enabling the Evaluation of Driver Physiology Via Vehicle Dynamics

\thanks{This
work received Horizon Europe funding via the ICT4CART
project under G.A. 768953. Content reflects only the authors’ view and the European Commission
is not responsible for any use that may be made of the information it contains. Pieces of this work are patent-protected \cite{pat1, pat2}.}
}

\author{\IEEEauthorblockN{Rodrigo Ordonez-Hurtado\IEEEauthorrefmark{1}}
\IEEEauthorblockA{\textit{Dublin Research Lab} \\
\textit{IBM Research Europe}\\
Dublin, Ireland \\
Rodrigo.Ordonez.Hurtado@ibm.com}
\and
\IEEEauthorblockN{Bo Wen\IEEEauthorrefmark{1}}
\IEEEauthorblockA{\textit{T.J. Watson Research Center} \\
\textit{IBM Research}\\
Yorktown Heights, USA \\
bwen@us.ibm.com}
\and
\IEEEauthorblockN{Nicholas Barra}
\IEEEauthorblockA{\textit{T.J. Watson Research Center} \\
\textit{IBM Research}\\
Yorktown Heights, USA \\
Nicholas.Barra@ibm.com}
\and
\IEEEauthorblockN{Ryan Vimba}
\IEEEauthorblockA{\textit{T.J. Watson Research Center} \\
\textit{IBM Research}\\
Yorktown Heights, USA \\
Ryan.Vimba@ibm.com}
\and

\linebreakand

\IEEEauthorblockN{Sergio Cabrero-Barros\IEEEauthorrefmark{2}}
\IEEEauthorblockA{\textit{Digital Media Department} \\
\textit{Fundaci\'{o}n Vicomtech}\\
Donostia / San Sebasti\'{a}n, Spain \\
scabrero@vicomtech.org}
\and
\IEEEauthorblockN{Sergiy Zhuk}
\IEEEauthorblockA{\textit{Dublin Research Lab} \\
\textit{IBM Research Europe}\\
Dublin, Ireland \\
Sergiy.Zhuk@ie.ibm.com}
\and
\IEEEauthorblockN{Jeffrey L. Rogers}
\IEEEauthorblockA{\textit{T.J. Watson Research Center} \\
\textit{IBM Research}\\
Yorktown Heights, USA \\
jeffrogers@us.ibm.com}

}

\maketitle

\begingroup\renewcommand\thefootnote{\IEEEauthorrefmark{1}}
\footnotetext{Equal contribution. \IEEEauthorrefmark{2}Contributed while affiliated to IBM Research.}
\endgroup

\begin{abstract}
Driving is a daily routine for many individuals across the globe. This paper presents the configuration and methodologies used to transform a vehicle into a connected ecosystem capable of assessing driver physiology. We integrated an array of commercial sensors from the automotive and digital health sectors along with driver inputs from the vehicle itself. This amalgamation of sensors allows for meticulous recording of the external conditions and driving maneuvers. These data streams are processed to extract key parameters, providing insights into driver behavior in relation to their external environment and illuminating vital physiological responses.
This innovative driver evaluation system holds the potential to amplify road safety. Moreover, when paired with data from conventional health settings, it may enhance early detection of health-related complications.
\end{abstract}


% ============================================
\section{Introduction}
\label{s_introduction}
% ============================================

% --------------------------------------------
\subsection{Motivation}
% --------------------------------------------
The COVID-19 pandemic starkly underscored the necessity for enhancing digital health solutions beyond the confines of traditional healthcare settings. Despite the growth in at-home medical services, the potential of in-vehicle environments as health data sources remain overlooked. As per an AAA survey \cite{AAA2020drivingstudy}, Americans clocked nearly 3 trillion miles and spent about 89 billion hours on the road from July 2019 to June 2020, averaging an hour-long, 30-mile trip each day. Given the substantial time people drive, this environment can offer unique insights into a driver's health status.

Driving necessitates a complex interplay of cognitive and physiological functions, including critical thinking, reaction times, stress responses, and muscle control. Any deterioration in these functions will yield observable alterations in how a vehicle is operated, with severe declines potentially leading to grave implications for  the driver and those around them. Evaluating these functions within other contexts, such as homes or clinics, can be challenging, inconvenient, or infrequent enough to detect minor fluctuations over time. At-home assessments may offer more regularity but often lack the necessary environmental stimuli to challenge these functions and demand high patient engagement.

Nonetheless, the recent strides in autonomous driving technology offer a wealth of tools for monitoring the vehicle's external environment, internal controls, and dynamic data. These tools enable the measurement of driver behaviors and assessment of cognitive and motor function performance, thereby shedding light on their underlying health condition.

These emerging data sources have the potential to considerably enhance the traditional clinic and home data streams, offering a more comprehensive view of an individual's health condition. Furthermore, drawing correlations between observations from various environments opens up exciting avenues for exploring questions, such as how a person's mental and physical state in one environment influences their behavior when transitioning into a different setting.

In this context, we propose a system that collects and processes vehicle dynamics data, which could pave the way for assessing driver physiology through vehicle dynamics. This approach effectively transforms a connected vehicle into a cost-effective health sensor.

    \begin{figure*}[t] %[!b]  % [ht] [thpb]
      \centering
        \includegraphics[width=0.8\linewidth]{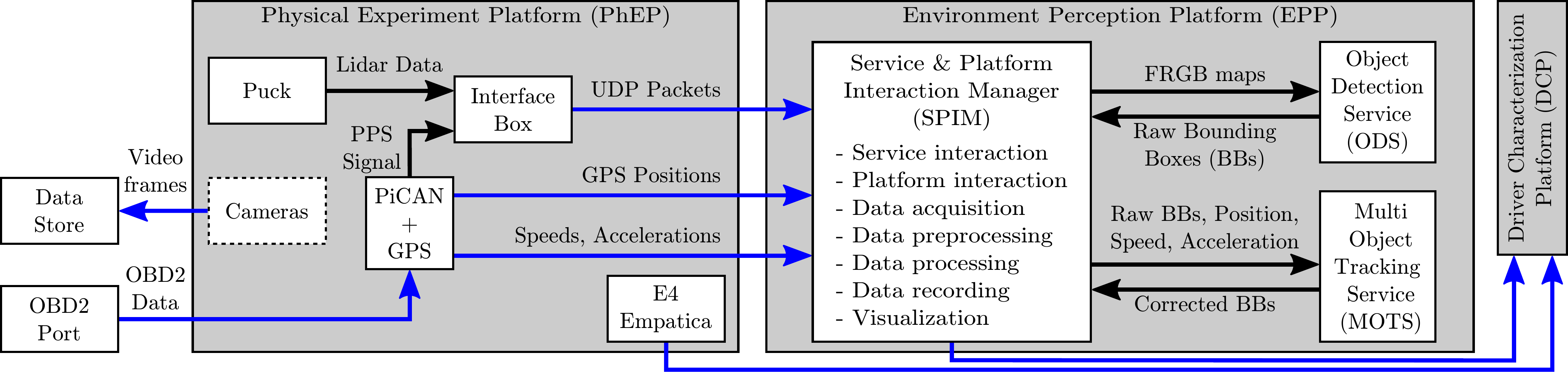}
      \caption{System Architecture of the proposed system, illustrating the main components of the Physical Experiment Platform and the Environment Perception Platform and their corresponding interactions. \textbf{REMARK:} Cameras are only used to support the labelling of training and validation data.}
      \label{f:system_architecture}
   \end{figure*}

% ============================================
\subsection{Proposed System}
\label{ss_proposed_study}
% ============================================
Our study aims to facilitate the analysis of potential relationships between driving behavior data and a driver's physiological signals. These relationships are instrumental in identifying relevant correlations for reliable estimation of either factor given the other. To achieve this, we designed and implemented HW/SW components \& tools for collecting, processing, estimating, and analyzing time-synchronized driving behavior and physiological data. Fig. \ref{f:system_architecture} shows the architecture of the proposed system, of which the main components/tools are:

1) The \underline{\textbf{Physical Experiment Platform (PhEP)}}, comprising the Wellness Utility Vehicle (WUV), instruments, and procedures to collect driver/vehicle data: a Data Collection Infrastructure (DCI) based on the Health Guardian \cite{hg-icdh2022} architecture, consisting of edge computing infrastructure, dynamic vehicle sensors, and driver cognition estimators; and operation of HW components (e.g., sensor calibration, device synchronization).

2) The \underline{\textbf{Environment Perception Platform (EPP)}}, comprising software modules to pre-process, aggregate, and exploit collected vehicle dynamics and remote sensing data (used for environment perception and scene identification): the Object Detection Service (ODS), to perform object detection on 2D maps from 3D Lidar data; the Multi-Object Tracking Service (MOTS), to enhance object detections obtained with the ODS; and the Service \& Platform Interaction Manager (SPIM), to integrate the EPP with other platforms and orchestrate the interaction with the ODS and MOTS services.

3) The \underline{\textbf{Driver Characterisation Platform (DCP)}}, comprising a driving parameter estimation service, a stress classification pipeline, and correlation analysis.

% --------------------------------------------
\subsection{Related work}
% --------------------------------------------
Lidar sensing, or 3D laser scanning, has become an industry standard for comprehending a vehicle's surroundings, thanks to the advent of autonomous vehicles and advancements in 3D object detection \cite{royo2019overview}. Complex-YOLO \cite{simony2018complex} is an efficient methodology for processing Lidar point clouds, building upon YOLO \cite{yolo1} and YOLOv2 \cite{yolo2}. It emphasizes real-time performance and heightened accuracy, utilizing solely point cloud data from a rotating Lidar sensor to create top-view maps with 2D detection bounding boxes. The later-developed Complexer-YOLO \cite{Complexer-YOLO} integrates this approach with semantic segmentation analysis of camera input, generating highly accurate 3D object detection and multi-target feature tracking.

Much research has aimed to quantify driver behavior to enhance autonomous driving systems and guarantee the safety of existing systems \cite{driving_behavior_3}. This work typically focuses on behaviors related to autonomous driving, such as drivers' attention patterns and reactions to various scenarios \cite{driving_behavior_1, driving_behavior_2}, using tools like eye-tracking and driver reaction data \cite{eye-tracking-research}. However, these studies primarily investigate the driver's environment rather than the driver. Our research takes a different approach, leveraging the environment to learn more about the driver and, in doing so, uncovers the potential for using the vehicle as a sensor of the driver's health condition.
% --------------------------------------------
\subsection{Document Structure}
% --------------------------------------------
Despite the successful design, implementation, integration, and validation of all subsystems illustrated in Fig. \ref{f:system_architecture}, this article primarily focuses on the PhEP/EPP due to space limitations. Detailed discussions regarding these systems can be found in Sections \ref{s:PhEP} and \ref{s:EPP}, respectively. A high-level overview of the DCP, along with selected outcomes, is given in Section \ref{s:Discussion}.

% ============================================
\section{The Physical Experiment Platform (PhEP)}
\label{s:PhEP}
% ============================================

\begin{figure}[ht]  % [thpb]
        \centering
        \includegraphics[width=0.8\linewidth]{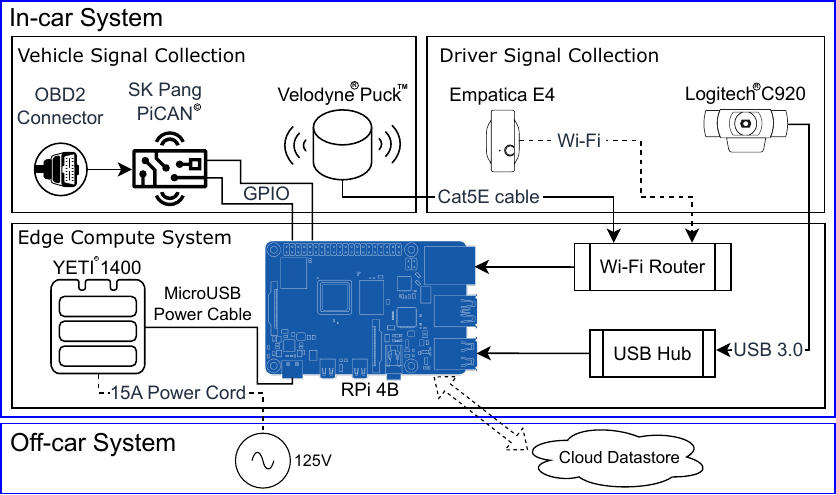}
        \caption{Wiring diagram of our system's components.}
        \label{fig:car-wire-diagram}
    \end{figure}

This platform comprises the proposed systems' hardware components, including the Wellness Utility Vehicle (WUV) and the supporting Data Collection Infrastructure (DCI). It consists of three major subsystems: the edge computing system, the vehicle signal collection system, and the driver signal collection system. We implemented the DCI with commercially available off-the-shelf components to ensure the solution quickly scales up at low cost. The locations of the DCI's main components are shown in Fig. \ref{fig:sensors}, their interconnections in Fig. \ref{fig:car-wire-diagram}, and their details are described in the following subsections.

% --------------------------------------------
\subsubsection{\textbf{Raspberry Pi}}
\label{ss:Lidar-sensor}
% --------------------------------------------
The central computing unit in our study is anchored by a Raspberry Pi 4 Model B. When we finished our experiments by early 2020, Raspberry Pi had just begun supporting computing clusters through Docker Swarm. It has evolved to accommodate microK8S-based clustering, facilitating a smooth expansion of computing capabilities.

% --------------------------------------------
\subsubsection{\textbf{SK Pang PiCAN}}
% --------------------------------------------

A PiCAN RPi HAT (Fig. \ref{fig:sensors}, item 1) with GPS, gyro/accelerometer, and CAN-Bus connector \cite{skpang} is connected to one of the RPi's GPIO ports via a ribbon cable. We can read key signals provided by the PiCAN directly from the RPi cluster and feed them into the data collection pipeline: some CAN bus signals measure the driver's operational actions (e.g., driving wheel angle and accelerator/brake pedal position), and onboard digital motion processor allows to use gyro/accelerometer to measure the vehicle's dynamics.
    
% --------------------------------------------
\subsubsection{\textbf{Lidar sensor}}
\label{ss:Lidar-sensor}
\begin{figure}[htbp]
        \centering
            \includegraphics[width=0.3\linewidth]{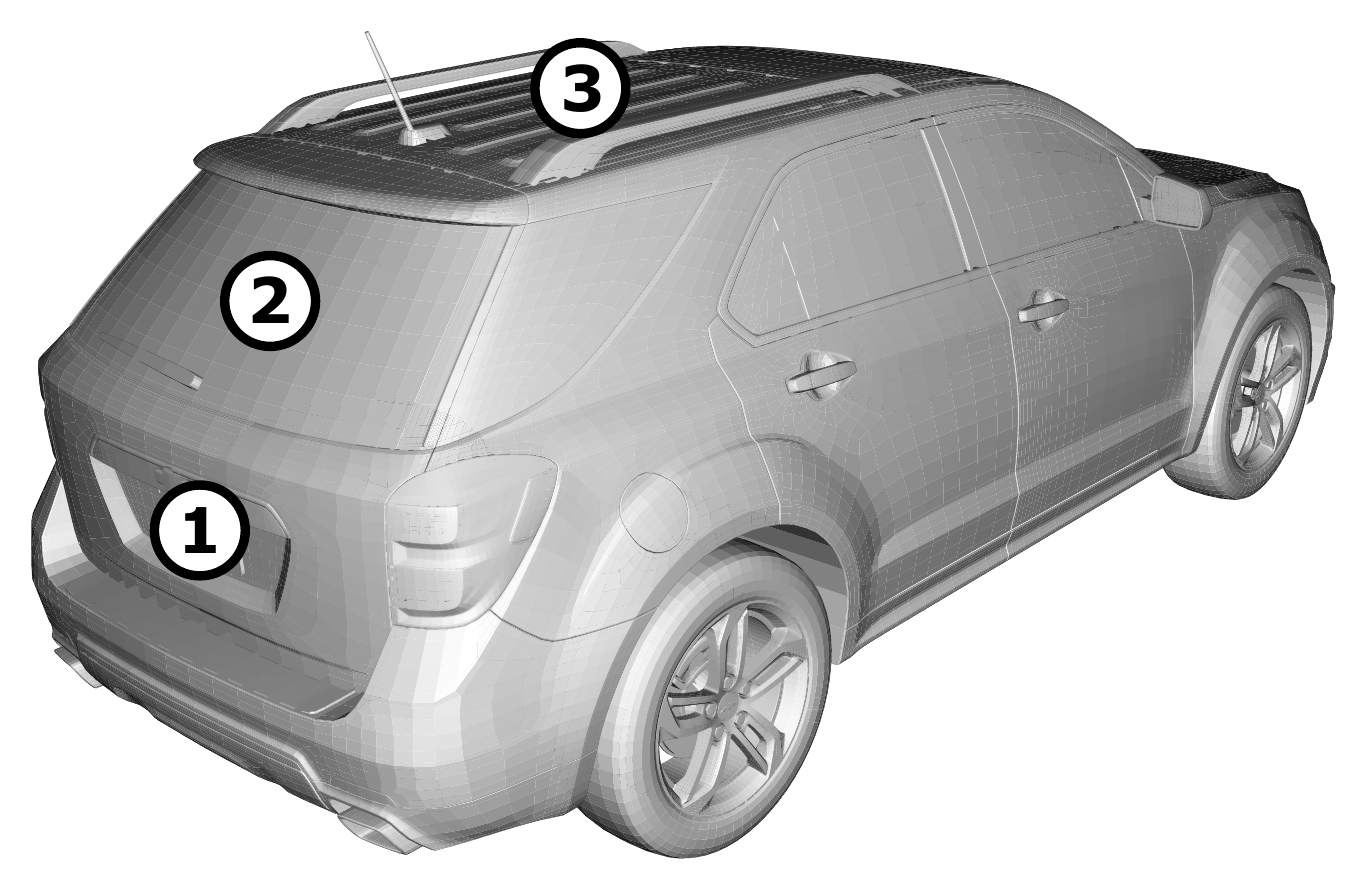}
            %\vfill
        \includegraphics[width=0.3\linewidth]{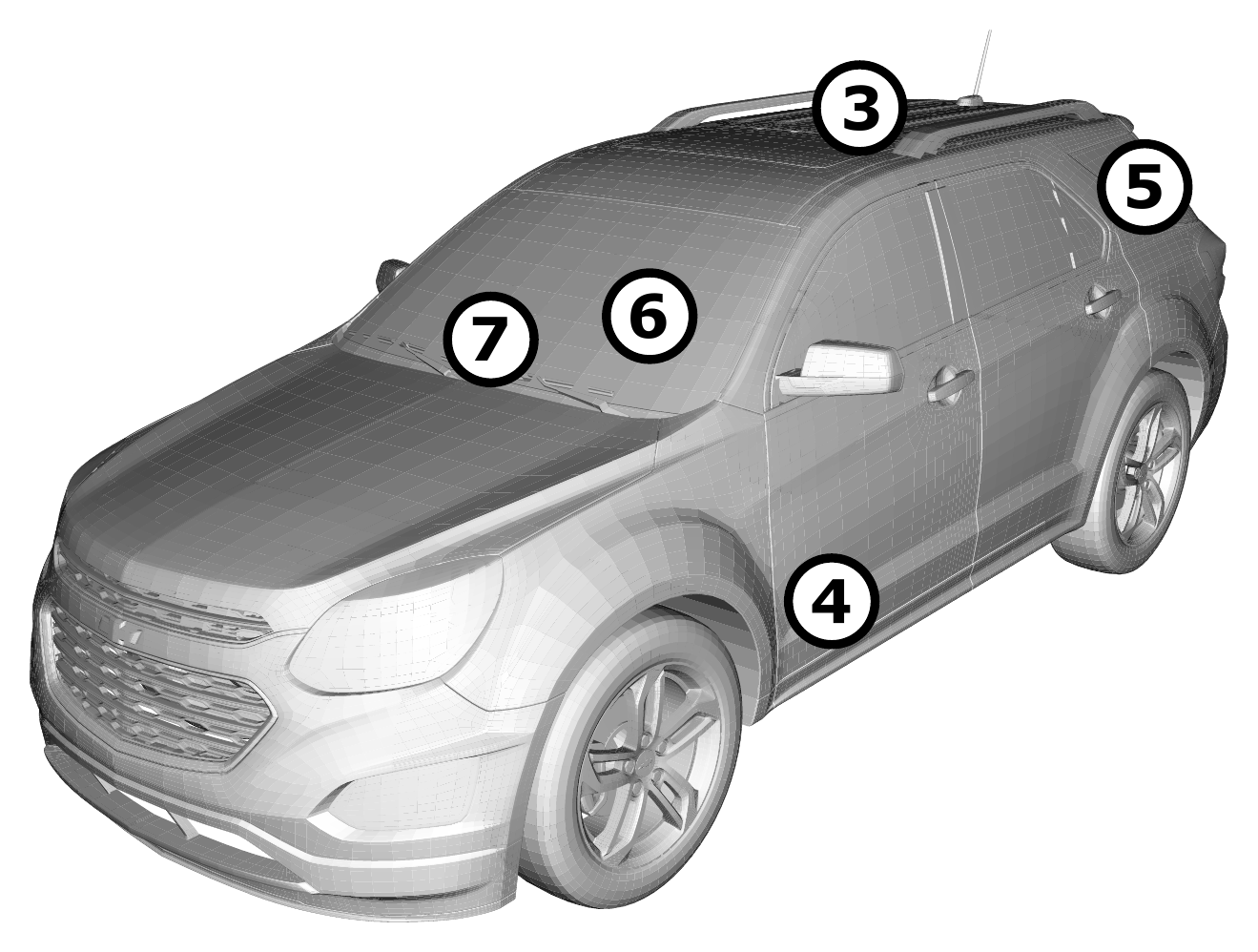}
    \caption{WUV's sensors: 1) RPi rack with PiCAN RPi HAT and RPi 4; 2) rear-view camera; 3) roof-mounted Lidar; 4) CAN bus cable connected to OBD2 port; 5) GPS antenna; 6) driver's face camera; 7) front-view camera.}
    \label{fig:sensors}
    \end{figure}

% --------------------------------------------
We mounted a Velodyne Puck Lidar sensor \cite{puck} on the WUV's roof (Fig. \ref{fig:sensors}, item 3) with a calibration process to minimize blind zones and mitigate ring mismatch effects. The Lidar's cable, running through the moonroof, connects to an Interface Box beneath the passenger seat for GPS stamp and time synchronization from the PiCAN. Lidar data packets are transmitted through the WUV's internal network, and the VeloView software \cite{veloview} on a networked laptop records the laser data, saving them as {\it .pcap} files.

% --------------------------------------------
\subsubsection{\textbf{Empatica E4 wristband}}
% --------------------------------------------

During the training phase, the wristband captured the driver's vital signs and subsequently uploaded them to the data storage after each session for offline analysis and model training. Alternatively, the wristband can be linked to the data collection pipeline through a smartphone, which serves as a bridge, converting the Bluetooth signal stream to Wi-Fi. This configuration allows for real-time analysis of the driver's vital signs.

% --------------------------------------------
\subsubsection{\textbf{Cameras}}
% --------------------------------------------
Video frames were utilized exclusively for data labeling, with no dependence on them for any analytics. We employed three Logitech\textsuperscript{\textregistered}\,C920 cameras - rear-view, front-view, and driver's face - all enhanced with the Kurokesu C920 enclosure kit \cite{kurokesu} for mounting CS-type fisheye lenses to broaden the view angle. Car electromagnetic noise posed challenges with long USB cables, causing camera connection instability. To counter this, we minimized the distance from each camera to the nearest USB data receiver and routed shielded cables from the hub to the RPi cluster, and ensured they do not lie close to any vehicle power transmission cables.
    
% ============================================
\section{The Environment Perception Platform (EPP)}
\label{s:EPP}
% ============================================

Visualizing driving as a computational task orchestrated within a driver's brain entails several steps: i) absorbing pertinent inputs from the environment, ii) constructing a mental model of the surroundings, iii) making informed decisions regarding vehicle navigation, and iv) executing appropriate actions based on these decisions. These input signals could come in various forms, such as visual cues like the distance to the vehicle ahead, auditory signals like honking, or tactile inputs like the feel of rumble strips.

To gain insight into how a driver's health condition influences their ability to manage this computational task, capturing and processing these stimulating input signals is crucial to ensure a precise understanding of the scene. This processing task is the primary function of the EPP platform, the components of which are illustrated in Fig. \ref{f:system_architecture} and detailed further in the following sections.
% --------------------------------------------
\subsection{The Object Detection Service (ODS)}
\label{ss_object_detection}
% --------------------------------------------

A Neural Network (NN) based ODS with GPU support was designed, implemented, and comprehensively validated using real-world Lidar data. The primary advancements associated with this service include i) creating an extensive database of labeled point clouds, ii) training various models for Lidar-based object detection, and iii) implementing the object detection functionality using a microservice architecture. The following subsections address these topics in detail.

\subsubsection{General Considerations}
\label{sss_ods_general_cons}
Advanced Lidar sensors for vehicular applications, like the HDL-64E sensor \cite{hdl64E}, often come with substantial costs, hindering their widespread use in real-world scenarios. Consequently, we opted for the more cost-efficient Puck unit (refer to Section \ref{ss:Lidar-sensor}). This decision influenced the selection of a detection model, necessitating compatibility with Puck data. At the outset of our study, there were no known benchmark suites designed specifically for the Puck sensor. Therefore, we created a Puck-like dataset by reducing the complexity of data from the KITTI Vision Benchmark Suite \cite{geiger2012kitti}. This dataset later played a pivotal role in developing a Puck dataset to train an object detection model tailored to the Puck sensor.

    \begin{figure}[ht]  % [thpb]
      \centering
        \includegraphics[width=0.8\linewidth]{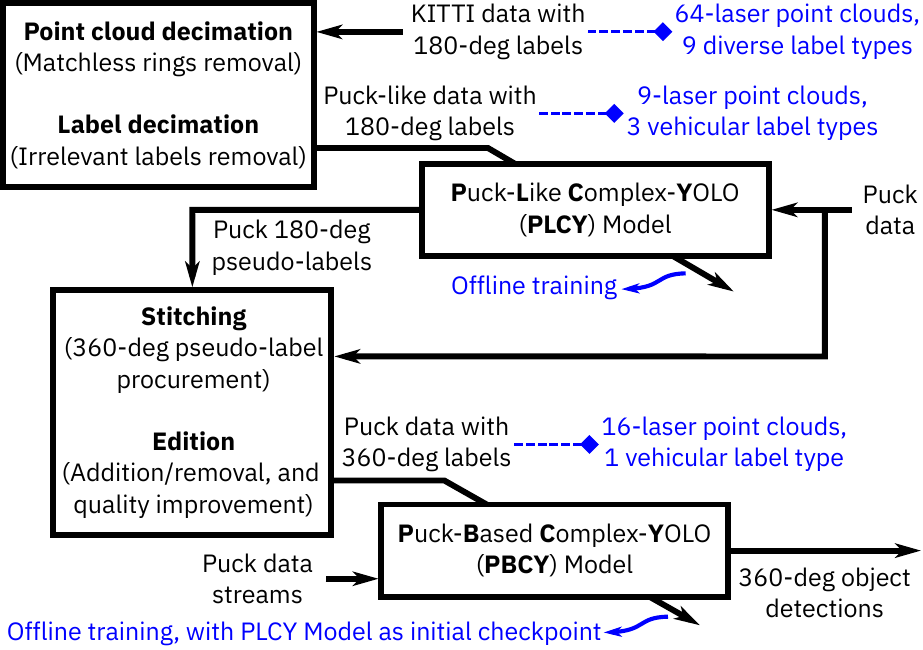}
      \caption{Pipeline to obtain full-azimuthal (360\degree) object detections.}
      \label{fig:plcy_pbcy_pipeline}
   \end{figure}

To mitigate the effects of reduced remote sensing data as a result of using the Puck, we shifted our focus to object detectors that do not directly rely on 3D point clouds but on 2D top-view FRGB maps instead and selected the Complex-YOLO network reported in \cite{simony2018complex} with three modifications:
\textbf{i)} train a first detection model using Puck-like data from KITTI data, namely a \textbf{P}uck-\textbf{L}ike \textbf{C}omplex-\textbf{Y}OLO (\textbf{PLCY}) model, for frontal object detection;
\textbf{ii)} design a 2-step workaround for the generation of full-azimuthal (360\degree) Puck labels; and
\textbf{iii)} ultimately train a detection model using the obtained full-azimuthal Puck labels, namely a \textbf{P}uck-\textbf{B}ased \textbf{C}omplex-\textbf{Y}OLO  (\textbf{PBCY}) model, for full-azimuthal object detection. The detailed rationale for the development of a Puck-fitted, full-azimuth object detector based on the previous modifications is shown in Fig. \ref{fig:plcy_pbcy_pipeline}, with all the involved steps addressed below.

\textbf{REMARK:} 360\degree\, detection is critical for diver characterization based on lane-changing models (e.g., \cite{MOBIL}), which use the information of leader/follower vehicles on surrounding lanes.
   
% ............................................
\subsubsection{Training of a PLCY Model for Frontal Object Detection}
\label{ss-PLCYM-training}
% ............................................

\begin{figure}[ht]  % [thpb]
      \centering
        \includegraphics[width=0.8\linewidth]{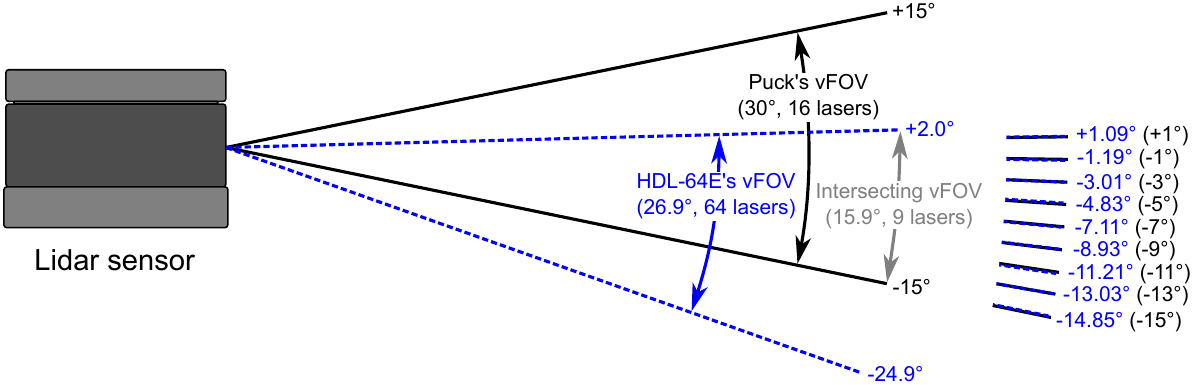}
      \caption{Comparison between the vertical field of view (vFOV) for the HDL-64E (blue) and the Puck (black) Lidar sensors.}
      \label{fig:HDL64vsVLP16}
   \end{figure}

The first step here is to obtain Puck-like data from KITTI data (which point clouds were collected with an HDL-64E sensor). As shown in Fig. \ref{fig:plcy_pbcy_pipeline}, this process involves:
\begin{itemize}
\item \textbf{Point cloud decimation}, as Puck and HDL-64E sensors have different vertical fields of view (vFOV) and different numbers of lasers with different inclination angles (see Fig. \ref{fig:HDL64vsVLP16}). Then, only 9 matching HDL-64E lasers in the intersecting vFOV are considered.
\item \textbf{Label decimation}, as KITTI has 9 diverse label types, of which we only need to focus on vehicular ones. Then, we only keep \textit{car}, \textit{van}, and \textit{truck} labels, and those with non-empty contents after the point cloud decimation.
\end{itemize}
   
Once the KITTI dataset was decimated, the resulting dataset was saved for the training of a PLCY model. The same point cloud preprocessing, loss function design, and training details provided in \cite{simony2018complex} are followed to achieve this. The core functionality implementation is loosely based on a publicly available GitHub repository \cite{complex_yolo_github}
with some required modifications and augmentations, such as added preprocessing functionality (including ground-surface removal and clustering).

The resulting Puck-like model trained with decimated KITTI data did not (expectedly) deliver favorable recall-precision curves (not included for space constraints) due to the emergence of insubstantial labels, which associated bounding boxes only covered a few datapoints. %
This resulted in a high rate of false negatives (i.e., recall figures are affected) and an increased potential for ``background noise'' to produce false positives (i.e., precision figures are affected).
However, the obtained PLCY model was suitable for creating a primal database of Puck pseudo-labels, as described next.

% ............................................
\subsubsection{PLCY-based Full-Azimuthal Object Detection}
\label{sss-PLCYM-for-360-detection}
% ............................................

\begin{figure}[ht]  % [thpb]
        \centering
        
        \includegraphics[width=0.8\linewidth] {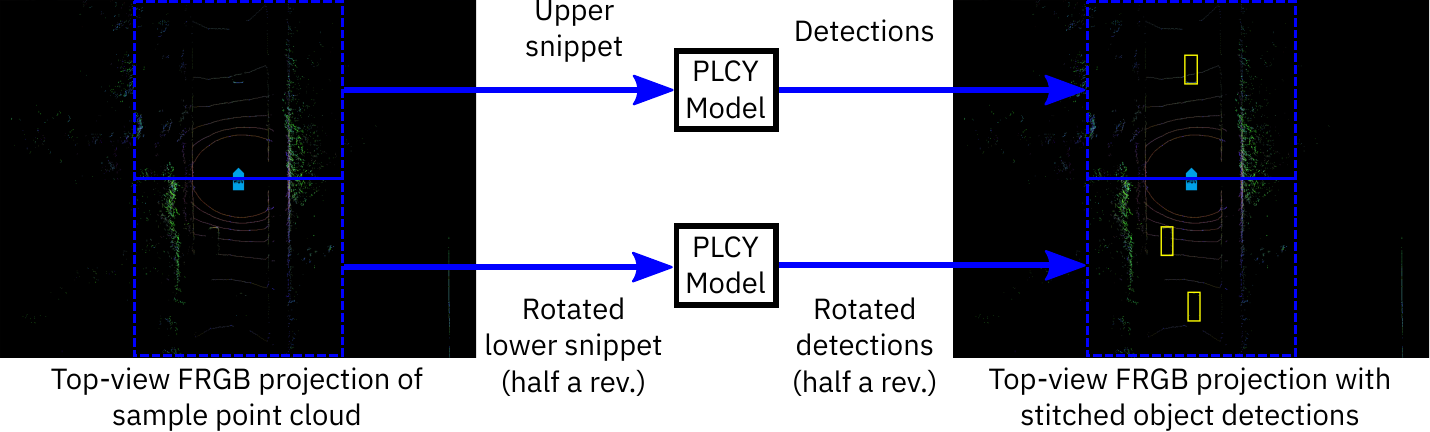}
        
        \caption{2-step full-azimuthal detection process using the PLCY model.}
        \label{fig:2-step-plcy}
    \end{figure}

To overcome the limitation of the PLCY model to only allow object detection in the WUV's front, We devised a 2-step detection process for each analyzed FRGB map. The actions required for this process are illustrated in Fig. \ref{fig:2-step-plcy} and explained below:

\begin{enumerate}
    \item[i)] Generate a top-view FRBG projection of the point cloud, and split it into upper and lower snippets corresponding to the WUV's front top-view and rear top-view halves.
    \item[ii)] Process each snippet in a separate detection step, which  requires a half-a-revolution rotation of the lower half.
    \item[iii)] Consolidate the obtained detections, which requires a half-a-revolution rotation of the lower half's detections.
\end{enumerate}

With the above procedure, the PLCY model can be used to create a database of Puck's full-azimuthal \textit{pseudo-labels} from point clouds collected with the Puck sensor during our field experiments in different traffic scenarios (urban and highway) and traffic conditions (free flow and congestion). We refer to them as ``pseudo-labels'' since they may not be entirely accurate but provide a readily available baseline of candidate labels that can be easily refined into accurate labels.
Over 1500 point clouds were collected from the test trips and processed with the PLCY model to obtain over 3000 pseudo-labels, which we turned into ground truth with a refining process that comprised relocation of boxes with shifted position, reorientation of boxes with shifted orientation, resizing of boxes with wrong size, deletion of boxes corresponding to false positives, and addition of missing boxes. As a result, we created an extensive database of over 1500 top-view FRGB maps with more than 4600 full-azimuthal labels.

% ............................................
\subsubsection{PBCY-based Full-Azimuthal Object Detection}
\label{ss-PBCYM-training}
% ............................................

We can now train a PBCY model using the Puck's ground truth database in Section \ref{sss-PLCYM-for-360-detection}, and following the same procedures described in Section \ref{ss-PLCYM-training} (but keeping ground surfaces since that proved to be beneficial).
The resulting precision-recall curves for the resulting trained PBCY model (not included here due to space limitations) exhibit an improved trade-off between the actual positive rate and the positive predictive value compared with those for the PLCY model.

% ............................................
\subsubsection{Microservice Implementation}
\label{ss-od-aas}
% ............................................

We encapsulated the PLCY/PBCY models in a microservice via a REST API deployed on a GPU-enabled server to allow remote access. The API call to get detections is invoked with top-view FRGB maps rather than with entire point clouds. In this way, the processing of massive point clouds (to obtain FRGB maps) can be performed locally in the WUV, and only lightweight FRGB maps need to be sent to the remote server hosting the ODS.
For offline experiments,
low latency is not required, and therefore a cloud deployment would suffice; if real-time performance is needed, a MEC deployment is advised.

% ............................................
\subsubsection{Comparative PLCY vs. PBCY Performance for Full-Azimuthal Object Detection}
\label{sss:PLCYM-vs-PBCYM}
% ............................................

\begin{figure}[htpb]  % [ht]  %  
        \centering
        
        \includegraphics[width=0.85\linewidth] {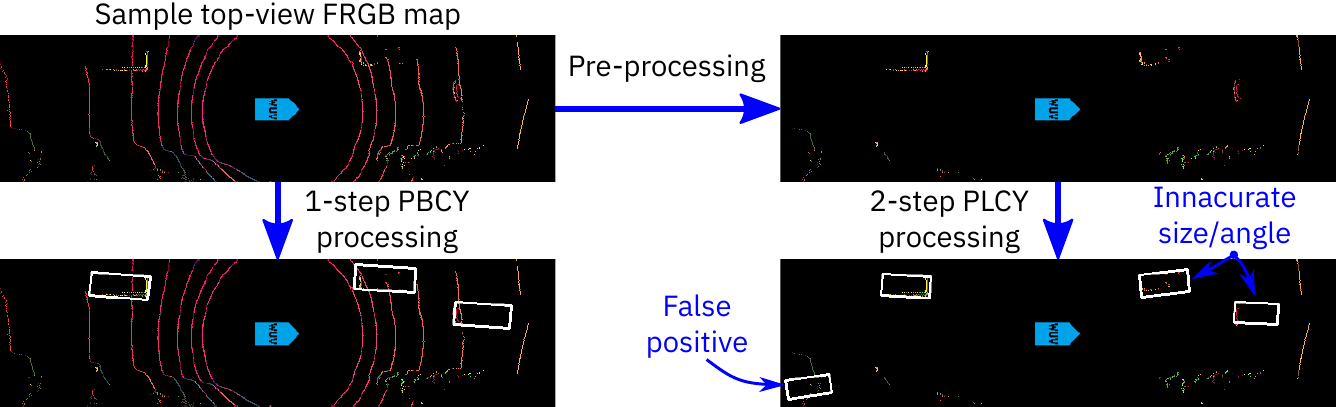}
        
        \caption{Comparison of PBCY/PLCY full-azimuthal object detection}
        \label{f:PLCYM-vs-PBCYM}
        
    \end{figure}

Figure \ref{f:PLCYM-vs-PBCYM} shows the detection outcome for a sample FRGB map processed by the PLCY and PBCY models. Beyond the additional actions required by the PLCY model to process a single frame (namely, pre-processing step and 2-step processing), Fig. \ref{f:PLCYM-vs-PBCYM} evidences the typical weaknesses of the PLCY model concerning the PBCY model: i) less-accurate size/angle of the positive detections, and ii) false detections due to remainders of the ground-surface removal process (not needed for the PBCY model).

% --------------------------------------------
\subsection{The Multi-Object Tracking Service (MOTS)}
\label{ss-object-tracking}
% --------------------------------------------

Continuous scene identification/understanding is crucial in our proposed system, which is evidenced by the already established need for monitoring different variables of (or w.r.t.) vehicles that surround the WUV: such monitoring must happen so that a driving model (usually employed to create traffic simulations) can be turned into an identification model that allows for the estimation of driving parameters, which ultimately leads to the characterization of the host driver. Thus, it is clear that our system must rely on two key features:

\begin{enumerate}
    \item[i)] \textbf{Accurate parameter estimation.} Driving models use nearby vehicles' data explicitly (e.g., positions) or indirectly (e.g., speeds derived from positions). Thus, such data must be estimated as accurately as possible.
    \item[ii)] \textbf{Accurate data association.} Calculating aggregated information such as relative speeds of ``objects of interest'' (e.g., a lane leader) requires tracking of detected objects between consecutive frames. Thus, object-ID matching must be done as accurately as possible to associate IDs with unique objects in different frames.
\end{enumerate}

% Concerning i),
Common Lidar-related issues such as measurement noise and partial object occlusion generally result in false positives/negatives. Also, the values of monitored variables in subsequent frames are independent noisy state measurements due to independent frame processing.
% Concerning ii),
On the other hand, lane-changing models generally require tracking of leader/ follower vehicles on current and target lanes, for which multiple objects must be monitored and tracked.
For these reasons (mainly), ``raw'' object detections from our ODS must be enhanced to prevent (or substantially mitigate) the effects of noisy/wrong/missing detections in downstream modules such as those in the Driving Characterization Platform. Therefore, a suitable Multi-Object Tracking (MOT) mechanism is needed.
Different methods to implement MOT systems with the described features have been reported in the specialized literature, being the most popular the Kalman Filter and Particle Filter techniques \cite{wang2020pointtracknet}. For illustration purposes, we have selected the former due to its proven benefits and ease of implementation (but other suitable approaches can be used).

% ............................................
\subsubsection{Kalman Filtering (KF) for Object Tracking}
\label{sss-KF-for-tracking}
% ............................................

KF is known for optimal state estimation from noisy signals \cite{kim2018-kf-introduction}, the primary technical advantage we exploit in our system. However, KF also enables other exciting features: i) the ability to deal with missing data via ``best-guess imputation'', e.g., to address missed detection of previously seen objects by calculating approximations with the transition equation; and ii) the option to track speeds if acceleration values are available (otherwise, a constant-velocity model is used).

    \begin{figure}[ht]  % [thpb]
      \centering
        \includegraphics[width=0.8\linewidth]{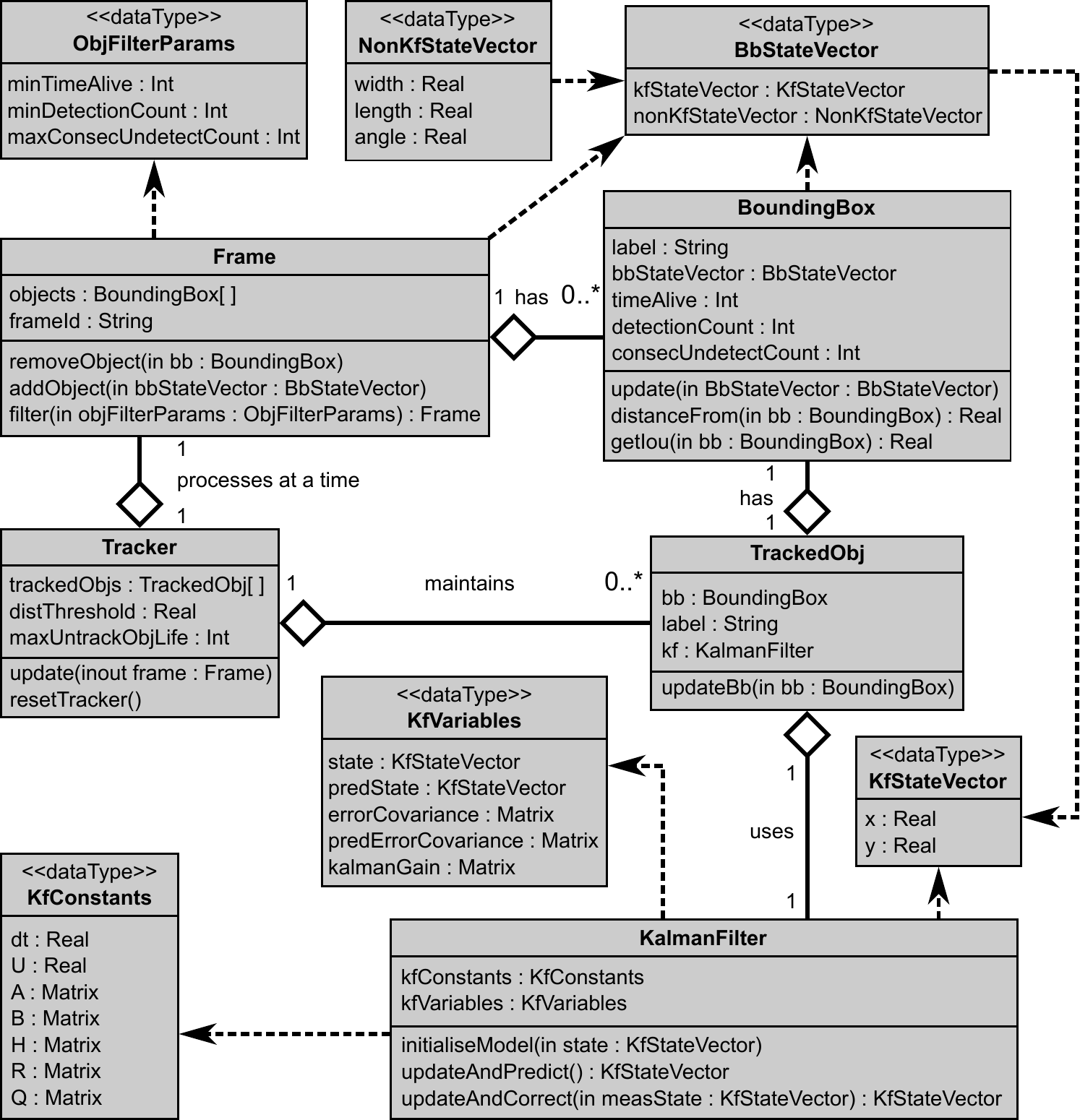}
      \caption{Simplified UML diagram of the Multi-Object Tracking Service.}
      \label{fig:UML-MOTS}
   \end{figure}

In our development, each tracked object is assigned a filter, and since the surrounding vehicles are considered 2D-moving objects (z-axis movement is neglected), a 2D-motion implementation of the filters with constant velocity is used \cite{kim2018-kf-introduction}.
A tracker instance is included so that \textbf{i)} tracked objects (and filters) can only be created/removed if specific conditions are met, and \textbf{ii)} the upcoming measurements are fed to the corresponding tracked objects. These management mechanisms adhere to the UML diagram presented in Fig. \ref{fig:UML-MOTS}: the {\it tracker} updates {frames} from the ODS, which contain multiple detections represented by {bounding boxes}. For this, the {\it tracker} i) creates new {\it tracked objects} by extending the {\it bounding boxes} of new detections with a {\it Kalman Filter}, and ii) modifies the attributes of existing {\it bounding boxes} via {\it prediction} (and {\it correction} if new measurements are available). The addition/removal of tracked objects is also controlled by policies with user-defined parameters (e.g., thresholds) that can be fine-tuned for specific scenarios. Our implementation is loosely based on a public GitHub repository \cite{kf_github}, 
with modifications/augmentations to, e.g., represent tracked objects as bounding boxes with shape and orientation, use Intersection-Over-Union-based ID assignment, and include {\it frame} and {\it bounding box} classes.
   
% ............................................
\subsubsection{Microservice Implementation}
% ............................................

Our KF-based MOT service was encapsulated as a microservice for its remote access via a REST API. The input of the API call to update a frame is the encoding of the associated detections as a {\it frame} object (i.e., a list of bounding boxes), and the corresponding response is the processed frame (i.e., a list of updated bounding boxes).

% ............................................
\subsubsection{Comparative Performance}
% ............................................

\begin{figure}[htpb]  % [ht]  %  
        \centering
    
        \includegraphics[width=0.8\columnwidth] {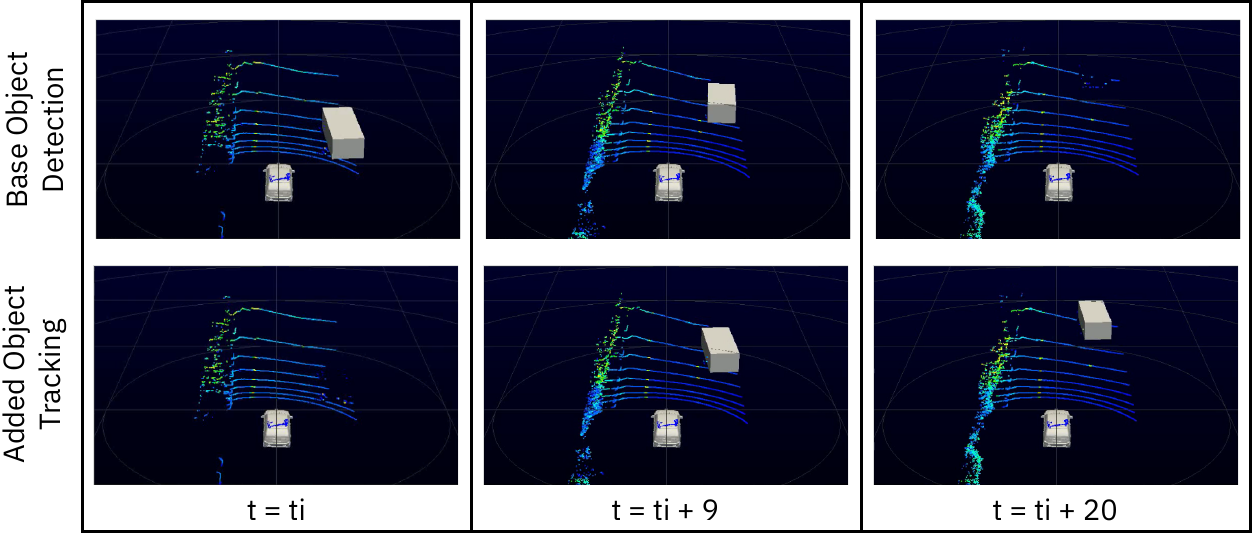}
        \caption{Examples of object detection with and without object tracking.}
        \label{fig:NoKF-vs-KF}
    \end{figure}

Figure \ref{fig:NoKF-vs-KF} illustrates the scene identification performance without and with object tracking. Some benefits of the latter addition are:
\textbf{i)} the omission of short-lived detections (at $ti$), which helps to filter out potential false positives;
\textbf{ii)} the inclusion of long-lived detections with KF corrections (at $ti+9$); and
\textbf{iii)} the inclusion of predicted detections during the temporary absence of associated detections (at $ti+20$), which helps to prevent false negatives.

% --------------------------------------------
\subsection{The Service \& Platform Interaction Manager (SPIM)}
\label{ss:SPIM}
% --------------------------------------------

The SPIM is the EPP's component in charge of the coordinated interaction with the PhEP, the services required for environmental perception, and the provision of supporting functionality to generate the driving-related data required by the Driver Characterization Platform. It comprises: i) intercommunication functionality to obtain vehicle-related data from the PhEP; ii) an orchestration module for the interaction with the ODS/MOTS services to obtain refined object detections; iii) functionality for leader/follower detection on current/next lanes; iv) data aggregation functionality to derive variables of interest (e.g., relative inter-vehicle speeds); and v) VeloView-based visualization.
Therefore, SPIM's submodules implement algorithms for point cloud rotation/truncation, FRGB map creation, ground surface removal (and point cloud clustering, if needed), and payload preparation for API calls, among others.
Our SPIM implementation was comprehensively validated with vehicle dynamics data collected via OBD2 port (stored in \textit{.hdf5} files) and ODS/MOTS outputs.

% ============================================
\section{Discussion}
\label{s:Discussion}

Our research presents a fresh viewpoint for digital health studies, recognizing driving as a significant daily activity for many. The developed PhEP and EPP platforms provide a holistic insight into a driver's \textit{input signal} and \textit{output action}, facilitating the deduction of driving behavior crucial for assessing cognitive health.

To exemplify our system's potential, we employed prototype Driver Characterization Platform that incorporates a Driving Parameter Estimation service based on the Intelligent Driver Model (IDM) \cite{IDM-2022}, an IBM Research asset for Stress Classification \cite{hao2018chrv}, and a correlation analysis via the Pearson correlation coefficient (PCC).

\begin{figure}[ht]  % [thpb]
      \centering
        \includegraphics[width=0.8\linewidth]{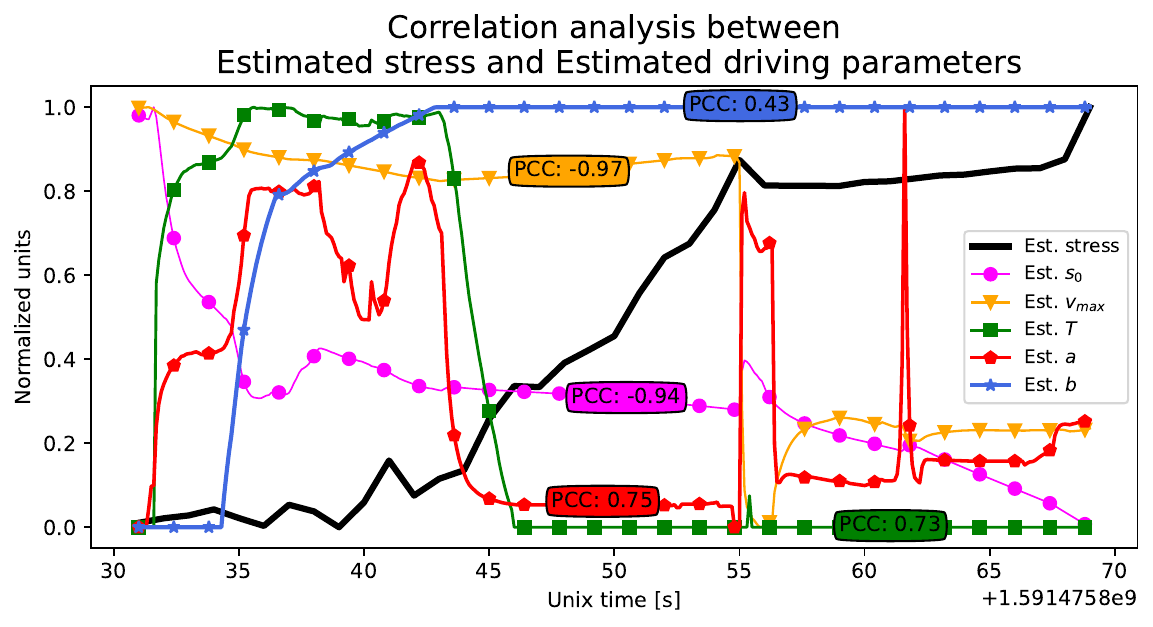}
      \caption{PCC values between estimations of IDM parameters and stress. [$s_0$: minimum spacing;
$v_{max}$: desired free-flow speed;
$T$: safe time headway;
$a$: maximum tolerated acceleration;
$b$: comfortable braking deceleration.]}
      \label{fig:Early-results}
    \end{figure}

\begin{figure}[ht]  % [thpb]
        \centering
    
        \subfloat[\label{sf:timestamp_53} Scene at 1591475853.]{\includegraphics[width=0.40\linewidth] {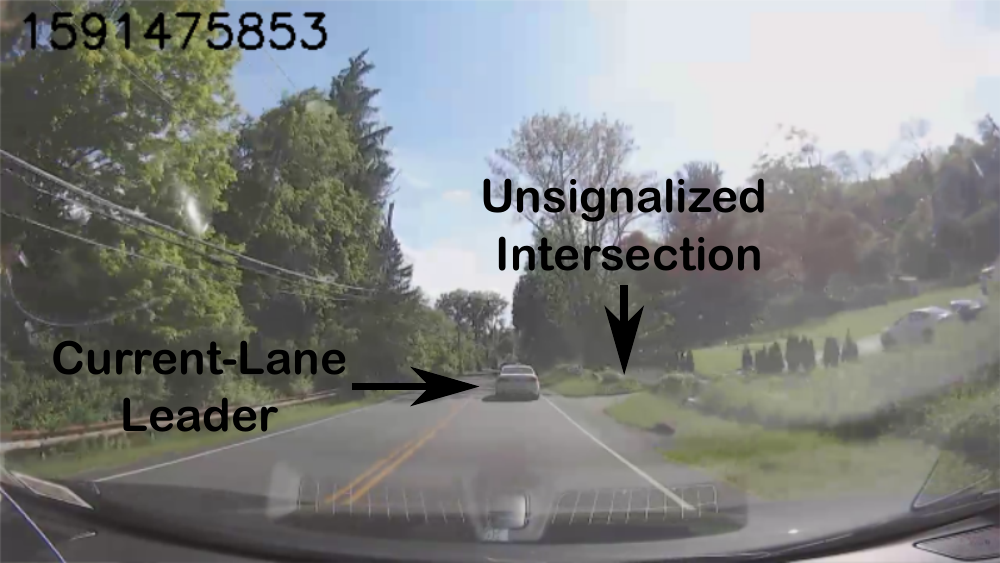}}\hspace{0.02\textwidth}%\hfill
        \subfloat[\label{sf:timestamp_54} Scene at 1591475854.]{\includegraphics[width=0.40\linewidth] {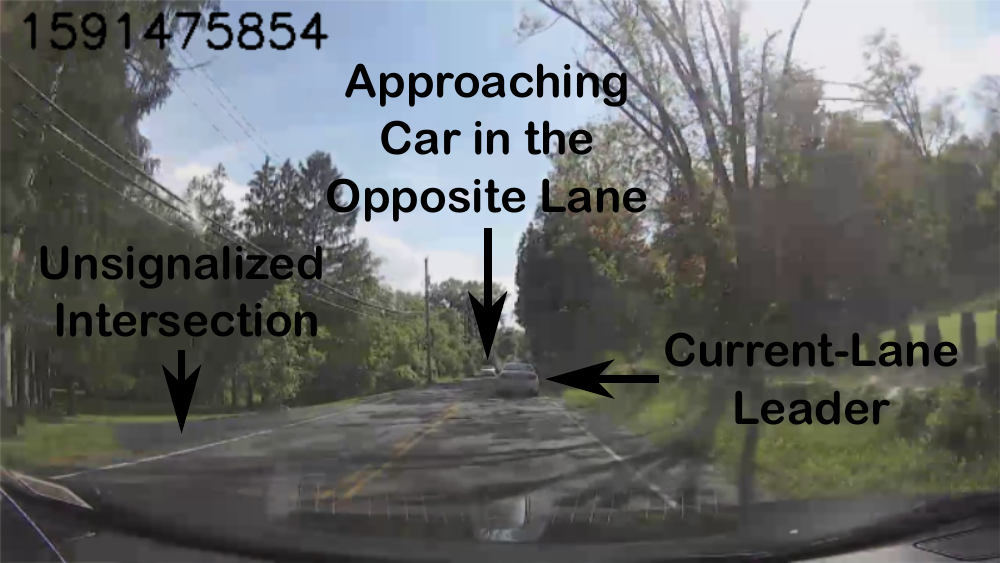}}\qquad
        
        \caption{Multiple tasks addressed by the driver toward timestamp 1591475855.}
        \label{fig:multitasking}
    \end{figure}

Figure \ref{fig:Early-results} displays the system's outcomes concerning the IDM's estimated driving parameters, the assessed driver's stress for a specific field experiment, and their corresponding correlation analysis. These findings highlight that: \textbf{i)} driving models utilizing vehicle dynamics can accurately characterize driving style, as estimated driving parameters reflect various facets of driver's actions, and \textbf{ii)} vehicle dynamics and physiological signals display inherent correlations due to the apparent interdependence between drivers' actions and their cognitive state. This is evident in Figs. \ref{fig:Early-results} and \ref{fig:multitasking} with:

\begin{enumerate}
\item The reported high and medium-high Pearson Correlation Coefficient (PCC) values for ${s_0, v_{max}}$ and ${T, a}$ respectively, calculated over the entire time window $[1591475830, 1591475870]$.
\item The concurrent abrupt changes in estimated parameters and stress signal around $t=1591475855$, which coincide with the driver facing a challenging situation, such as approaching two uncontrolled junctions while maintaining a safe distance from the vehicle ahead.
\end{enumerate}


These observations present potential applications such as \textit{anomaly detection} relative to a known driving profile, and \textit{virtual health sensor} to estimate cognitive state without physiological sensors. Future work includes refining these applications, assessing advanced 2D object detection approaches, examining lane-changing driving models (e.g., \cite{MOBIL}) for additional driving parameters like politeness factor, and studying other physiological responses like attention drift.

% ============================================

\section*{Acknowledgment}
Thanks to Dr. J. Monteil and Dr. N. Hinds for their key insights, and Dr. T. Hao for the guidance on stress classification.

\bibliographystyle{IEEEtran}
\bibliography{references}

\begin{thebibliography}{10}
\providecommand{\url}[1]{#1}
\csname url@samestyle\endcsname
\providecommand{\newblock}{\relax}
\providecommand{\bibinfo}[2]{#2}
\providecommand{\BIBentrySTDinterwordspacing}{\spaceskip=0pt\relax}
\providecommand{\BIBentryALTinterwordstretchfactor}{4}
\providecommand{\BIBentryALTinterwordspacing}{\spaceskip=\fontdimen2\font plus
\BIBentryALTinterwordstretchfactor\fontdimen3\font minus
  \fontdimen4\font\relax}
\providecommand{\BIBforeignlanguage}[2]{{%
\expandafter\ifx\csname l@#1\endcsname\relax
\typeout{** WARNING: IEEEtran.bst: No hyphenation pattern has been}%
\typeout{** loaded for the language `#1'. Using the pattern for}%
\typeout{** the default language instead.}%
\else
\language=\csname l@#1\endcsname
\fi
#2}}
\providecommand{\BIBdecl}{\relax}
\BIBdecl

\bibitem{pat1}
J.~Monteil, Y.~Lassoued, S.~C. Barros, R.~H. Ordonez-Hurtado, M.~Mevissen,
  S.~Zhuk, N.~Hinds, B.~Wen, and J.~Rogers, ``Mitigating risk behaviors,'' U.S.
  Patent 20\,210\,107\,501, April 15, 2021.

\bibitem{pat2}
E.~Bilal, B.~Wen, N.~A. Barra, J.~L. Rogers, B.~Dang, and T.~Hao, ``Assessing
  driver cognitive state,'' U.S. Patent 20\,220\,363\,264, November 17, 2022.

\bibitem{AAA2020drivingstudy}
{AAA Foundation for Traffic Safety}, ``{New American Driving Survey: Updated
  Methodology and Results from July 2019 to June 2020},'' AAA Foundation for
  Traffic Safety, Washington, DC 20005, Tech. Rep., Apr 2021.

\bibitem{hg-icdh2022}
B.~Wen, V.~S. Siu, I.~Buleje, K.~Y. Hsieh, T.~Itoh, L.~Zimmerli, N.~Hinds,
  E.~Eyigöz, B.~Dang, S.~von Cavallar, and J.~L. Rogers, ``{Health Guardian
  Platform: A technology stack to accelerate discovery in Digital Health
  research},'' in \emph{2022 IEEE International Conference on Digital Health
  (ICDH)}, 2022, pp. 40--46.

\bibitem{royo2019overview}
S.~Royo and M.~Ballesta-Garcia, ``An overview of lidar imaging systems for
  autonomous vehicles,'' \emph{Applied sciences}, vol.~9, no.~19, p. 4093,
  2019.

\bibitem{simony2018complex}
M.~Simon, S.~Milz, K.~Amende, and H.-M. Gross, ``{Complex-YOLO: An
  Euler-Region-Proposal for Real-Time 3D Object Detection on Point Clouds},''
  in \emph{Computer Vision -- ECCV 2018 Workshops}, L.~Leal-Taix{\'e} and
  S.~Roth, Eds.\hskip 1em plus 0.5em minus 0.4em\relax Cham: Springer
  International Publishing, 2019, pp. 197--209.

\bibitem{yolo1}
J.~Redmon, S.~Divvala, R.~Girshick, and A.~Farhadi, ``{You Only Look Once:
  Unified, Real-Time Object Detection},'' in \emph{Proceedings of the IEEE
  conference on computer vision and pattern recognition}, 2016, pp. 779--788.

\bibitem{yolo2}
J.~Redmon and A.~Farhadi, ``{YOLO9000: Better, Faster, Stronger},'' in
  \emph{Proceedings of the IEEE conference on computer vision and pattern
  recognition}, 2017, pp. 7263--7271.

\bibitem{Complexer-YOLO}
M.~Simon, K.~Amende, A.~Kraus, J.~Honer, T.~Sämann, H.~Kaulbersch, S.~Milz,
  and H.~M. Gross, ``{Complexer-YOLO: Real-Time 3D Object Detection and
  Tracking on Semantic Point Clouds},'' in \emph{2019 IEEE/CVF Conference on
  Computer Vision and Pattern Recognition Workshops (CVPRW)}, 2019, pp.
  1190--1199.

\bibitem{driving_behavior_3}
F.~Codevilla, E.~Santana, A.~M. Lopez, and A.~Gaidon, ``Exploring the
  limitations of behavior cloning for autonomous driving,'' in
  \emph{Proceedings of the IEEE/CVF International Conference on Computer Vision
  (ICCV)}, October 2019.

\bibitem{driving_behavior_1}
M.~Bojarski, D.~Del~Testa, D.~Dworakowski, B.~Firner, B.~Flepp, P.~Goyal, L.~D.
  Jackel, M.~Monfort, U.~Muller, J.~Zhang, X.~Zhang, J.~Zhao, and K.~Zieba,
  ``End to end learning for self-driving cars,'' \emph{arXiv preprint
  arXiv:1604.07316}, 2016.

\bibitem{driving_behavior_2}
F.~Codevilla, M.~Miiller, A.~L\'{o}pez, V.~Koltun, and A.~Dosovitskiy,
  ``End-to-end driving via conditional imitation learning,'' in
  \emph{Proceedings of IEEE International Conference on Robotics and Automation
  (ICRA)}.\hskip 1em plus 0.5em minus 0.4em\relax IEEE, 2018, pp. 4693--4700.

\bibitem{eye-tracking-research}
\BIBentryALTinterwordspacing
C.~Liu, Y.~Chen, L.~Tai, H.~Ye, M.~Liu, and B.~E. Shi, ``A gaze model improves
  autonomous driving,'' in \emph{Proceedings of the 11th ACM Symposium on Eye
  Tracking Research \& Applications}, ser. ETRA '19.\hskip 1em plus 0.5em minus
  0.4em\relax New York, NY, USA: Association for Computing Machinery, 2019.
  [Online]. Available: \url{https://doi.org/10.1145/3314111.3319846}
\BIBentrySTDinterwordspacing

\bibitem{skpang}
{SK Pang}, ``{PiCAN with GPS + Gyro +Accelerometer CAN-Bus for Raspberry Pi},''
  \url{https://www.skpang.co.uk/products/pican-with-gps-gyro-accelerometer-can-bus-for-raspberry-pi-3},
  {A}ccessed: 2023-06-03.

\bibitem{puck}
{Velodyne Lidar}\textregistered, ``{Puck lidar sensor},''
  \url{https://velodynelidar.com/products/puck/}, {A}ccessed: 2023-06-03.

\bibitem{veloview}
{ParaView}, ``{VeloView: The Velodyne Lidar Viewer based on Paraview Lidar},''
  \url{https://www.paraview.org/veloview/}, {A}ccessed: 2023-06-03.

\bibitem{kurokesu}
{Kurokesu}, ``{C920/C922/C930 enclosure kit for CS-type lens mk2},''
  \url{https://www.kurokesu.com/shop/diy_kits/C920_REWORK_KIT2}, {A}ccessed:
  2023-06-03.

\bibitem{hdl64E}
{Velodyne Lidar}\textregistered, ``{High Definitian Lidar HDL-64E - Data
  Sheet},''
  \url{https://hypertech.co.il/wp-content/uploads/2015/12/HDL-64E-Data-Sheet.pdf},
  {A}ccessed: 2023-06-03.

\bibitem{geiger2012kitti}
A.~Geiger, P.~Lenz, and R.~Urtasun, ``Are we ready for autonomous driving? {The
  KITTI Vision Benchmark Suite},'' in \emph{Proc. IEEE Comput. Soc. Conf.
  Comput. Vis. Pattern Recognit.}\hskip 1em plus 0.5em minus 0.4em\relax IEEE,
  2012, pp. 3354--3361.

\bibitem{MOBIL}
A.~Kesting, M.~Treiber, and D.~Helbing, ``General lane-changing model mobil for
  car-following models,'' \emph{Transportation Research Record}, vol. 1999,
  no.~1, pp. 86--94, 2007.

\bibitem{complex_yolo_github}
{Fudongxu}, ``{Complex-YOLO implementation in tensorflow},''
  \url{https://github.com/wwooo/tensorflow_complex_yolo}, {A}ccessed:
  2023-06-03.

\bibitem{wang2020pointtracknet}
S.~Wang, Y.~Sun, C.~Liu, and M.~Liu, ``{PointTracknet: An End-to-End Network
  for 3-D Object Detection and Tracking from Point Clouds},'' \emph{IEEE
  Robotics and Automation Letters}, vol.~5, no.~2, pp. 3206--3212, 2020.

\bibitem{kim2018-kf-introduction}
Y.~Kim and H.~Bang, ``{Introduction to Kalman Filter and its Applications},''
  \emph{Introduction and Implementations of the Kalman Filter}, vol.~1, pp.
  1--16, 2018.

\bibitem{kf_github}
A.~Sharma, ``{Multi Object Tracking with Kalman-Filter},''
  \url{https://github.com/mabhisharma/Multi-Object-Tracking-with-Kalman-Filter},
  {A}ccessed: 2023-06-03.

\bibitem{IDM-2022}
S.~Albeaik, A.~Bayen, M.~T. Chiri, X.~Gong, A.~Hayat, N.~Kardous, A.~Keimer,
  S.~T. McQuade, B.~Piccoli, and Y.~You, ``{Limitations and Improvements of the
  Intelligent Driver Model (IDM)},'' \emph{SIAM Journal on Applied Dynamical
  Systems}, vol.~21, no.~3, pp. 1862--1892, 2022.

\bibitem{hao2018chrv}
T.~Hao, H.~Chang, M.~Ball, K.~Lin, and X.~Zhu, ``{cHRV Uncovering Daily Stress
  Dynamics Using Bio-Signal from Consumer Wearables},'' in \emph{Proceedings of
  the 16th World Congress on Medical and Health Informatics}, vol. 245, 2018,
  p.~98.

\end{thebibliography}

\end{document}